\documentclass[12pt]{article}
\usepackage{epsfig}
\usepackage{graphicx}
 \usepackage{latexsym,amsmath,tabularx,amssymb,bm}

\newcommand{\be}{\begin{equation}}
\newcommand{\ee}{\end{equation}}
\newcommand{\bea}{\begin{eqnarray}}
\newcommand{\eea}{\end{eqnarray}}

\newcommand{\ep}{\epsilon}

\def\simge{\mathrel{%
   \rlap{\raise 0.511ex \hbox{$>$}}{\lower 0.511ex \hbox{$\sim$}}}}
\def\simle{\mathrel{
   \rlap{\raise 0.511ex \hbox{$<$}}{\lower 0.511ex \hbox{$\sim$}}}}
\def\bigs{\mathrel{
   \rlap{\raise 0.531ex \hbox{$>$}}{\lower 0.531ex \hbox{$<$}}}}

\def\del{\partial}                              

\begin{document}

\title{\large \bf Ward identities for amplitudes with reggeized gluons}
\author{
{\large J.~Bartels$^{1,2}$, L.~N. Lipatov$^{1,3}$ and G.~P.~Vacca$^{4}$} \bigskip \\
\small {\it $^1$~II. Institute of  Theoretical Physics, Hamburg University, Germany} \\
\small {\it $^2$~ Departamento de F\'{i}sica, Universidad T\'{e}cnica Federico Santa Mar\'{i}a, Valparaiso, Chile}\\
\small {\it $^3$~St. Petersburg Nuclear Physics Institute, Russia}\\
\small {\it $^4$~INFN - Sezione di Bologna,  Bologna, Italy}}
\date{}
\maketitle

\vspace{-9cm}
\begin{flushright}
DESY-12-071
\end{flushright}
\vspace{7cm}

\abstract{
\noindent
Starting from the effective action of high energy QCD we derive Ward identities  for Green's functions of reggeized gluons. They follow from the gauge invariance of the effective action, and 
allow to derive new representations of amplitudes containing physical particles as well as 
reggeized gluons. We explicitly demonstrate their validity for the BFKL kernel, and we present a new 
derivation of the kernel.}

\section{Generalized Ward Identities}
Gauge invariance has been the guide to construct, within QCD, an effective action~\cite{effact,Lipatov:1996ts}  
which introduces the fields of reggeized gluons and describes the high energy behavior 
of QCD. It automatically leads to the construction of gauge invariant amplitudes and Green's functions of reggeized gluon and physical particles.
The effective action generates a set of extended Feynman rules~\cite{feynruleffact} with 
interactions which are local in rapidity and which may be used to 
compute amplitudes involving reggeized gluons. 
Since reggeized gluons are off shell and belong to unphysical 
polarizations\footnote{Really the reggeized gluons can be considered as 
gauge invariant states having the physical polarizations in the 
crossing channel $t$ and lying on the Regge trajectory.}, it is important to investigate symmetry properties derived from gauge invariance. As in normal QCD, gauge symmetry plays an important role in doing  explicit calculations.

In this paper we extend the BRST invariance of QCD to Green's functions of reggeized gluons
and derive a generalized set of Ward identities (Section 1). We find it convenient to first recapitulate  
a few identities for amplitudes in normal QCD. We then extend these identities to 
the Green's functions and amplitudes derived from the effective action case. 
In the second part of our paper (Section 2) we demonstrate, as a first  application, the 
validity of these Ward identities for the 4-point function of four reggeized gluons, and we 
give a new derivation of the BFKL kernel.  

\subsection{QCD}
In this section we remind which kind of simple QED-like Ward identities we may expect for a general QCD scattering amplitude.

QCD is a theory with $SU(N_c)$ Yang-Mills gluon fields coupled to quarks.
This theory has a unitary S-matrix  presented in terms of transversely polarized colored gluon and quark fields as on shell asymptotic states, 
even if these are not the truly physical states as hadrons and mesons of the confining phase.

Let us denote a physical transverse gluon polarization vector ($\lambda=1,2$ in four dimensions) $\ep_{(\lambda)}^{ \mu}(k)$.
The corresponding states satisfy the Lorentz condition $k_\mu \ep_{(\lambda)}^{ \mu}(k)=0$. The additional solutions of this equation with the longitudinal polarization $\ep_L^\mu(k)=c(k) k_\mu$ are decoupled from the physical states. In other words the scattering amplitude with such longitudinal gluons is zero.

In the following  we shall just consider amplitudes with external physical on shell quarks and gluons and possibly one or more longitudinally polarized gluons. They satisfy a tower of Ward-like identities. We can interpret them as saying that longitudinal polarization states ($\ep^L$) decouple from physical polarization states ($\ep^\lambda$). 
In practice one has associated to the on shell physical scattering amplitude
\be
M=\ep_1^{\mu_1}(q1)\ep_2^{\mu_2}(q_2)\cdots \ep_n^{\mu_n}(q_n) M_{\mu_1 \mu_2 \cdots \mu_n}(q_1,q_2, \cdots , q_n)
\ee
the tower of identities
\bea
&{}&q_1^{\mu_1}\ep_2^{\mu_2}(q_2)\ep_3^{\mu_3}(q_3)\cdots \ep_n^{\mu_n}(q_n) M_{\mu_1 \mu_2 \cdots \mu_n}(q_1,q_2, q_3,\cdots , q_{n-1},q_n)=0 \nonumber\\
&{}& \cdots \nonumber\\
&{}&q_1^{\mu_1} q_2^{\mu_2}\ep_3^{\mu_3}(q_3)\cdots \ep_{n-1}^{\mu_{n-1}}(q_{n-1})\ep_n^{\mu_n}(q_n) M_{\mu_1 \mu_2 \cdots \mu_n}(q_1,q_2, q_3,\cdots ,q_{n-1}, q_n)=0 \nonumber\\
&{}& \cdots \nonumber\\
&{}&q_1^{\mu_1} q_2^{\mu_2}\cdots q_{n-1}^{\mu_{n-1} }\ep_n^{\mu_n}(q_n) M_{\mu_1 \mu_2 \cdots \mu_n}(q_1,q_2, q_3,\cdots , q_{n-1},q_n)=0 \nonumber\\
&{}& \cdots \nonumber\\
&{}&q_1^{\mu_1} q_2^{\mu_2}\cdots q_{n-1}^{\mu_{n-1} }q_n^{\mu_n} M_{\mu_1 \mu_2 \cdots \mu_n}(q_1,q_2, q_3,\cdots , q_{n-1},q_n)=0.
\label{qcd_id}
\eea
Any single or multiple contraction with longitudinal polarizations gives zero if all the other lines are contracted with physical polarization vectors and are on shell. This is a consequence of the fact that QCD is a gauge invariant theory.

Let us now recall how to prove these identities\footnote{This proof is not original and can be found in the literature.}.
It is convenient to use the global BRST symmetry of the gauge fixed QCD action.
Let us start from the QCD lagrangian density with the so called general Lorentz gauge fixing (Lorentz invariant) and the associated ghost terms included (restricting to the gluon sector (field $v^a_\mu$) since the quark sector is trivial in our analysis)
\be
{\cal L}(v_\mu)=-\frac{1}{4}F_{\mu\nu}^a F^{a\mu\nu} -  v^a_\mu \partial^\mu B^a -\frac{\xi}{2} B^a B^a+\partial^\mu \bar{c}^a (D_\mu c)^a \,,
\label{lag_QCD}
 \ee
 where $B^a$ is an auxiliary field which satisfies the equation of motion $B^a=\frac{1}{\xi} \partial^\mu v^a_\mu$ and $D_\mu$ is the usual covariant derivative.
 One can therefore write the conjugated momenta
 \be
 \pi_i^a=F^a_{0i} \quad , \quad \pi_B^a=-v^a_0 \quad , \quad \pi^a_c=-\partial_0 \bar{c}^a \quad , \quad \pi^a_{\bar{c}}=(D_0 c)^a\,.
 \ee
 The global BRST symmetry is defined by
 \be
 \delta_\ep v^a_\mu=\ep (D_\mu c)^a\quad , \quad \delta_\ep c^a = -\frac{1}{2}\ep g f^{ade}c^d c^e
 \quad \delta_\ep \bar{c}^a= \ep B^a\quad \delta_\ep B^a =0 \,,
 \ee
 where $\ep$ is the infinitesimal Grassmann parameter of the transformation. In terms of the momenta one can immediately write, using the N\"other theorem, the conserved BRST charge 
 \be
 Q=\int d^3 x \left[ \pi^a_i (D_i \,c)^a -\frac{g}{2} f^{ade} \pi^a_c c^d c^e+\pi^a_{\bar{c}} B^a \right] \,,
 \ee
 which, after quantization, generates the quantum BRST transformation.

 The BRST transformation is nilpotent and therefore $Q^2=0$.
 We remind that the "large" Hilbert space can contain states with negative norm. There are, at fixed color,  six different asymptotic states: 
 $v_\mu^{a\, (transverse)}$ are the two states with physical polarizations, then there are other two gluons states, one of them described by the auxiliary field $B^a$, and finally there are the $c^a$ ghost and the $\bar{c}^a$ antighost fields.
 
 The physical states are the transverse polarized gluons which are annihilated by the BRST charge $Q$.
 The subspace of the large Hilbert space which belongs to the kernel of $Q$ contains the physical states plus the zero norm states ($|B\rangle$ and $|c \rangle$ ). The physical Hilbert space is the quotient of such a space with respect to the subspace of zero norm states, i.e.
 ${\cal H}_{phys}={\rm Cohomology(Q)}={\rm ker}\, Q / {\rm im}\, Q $.
 From the BRST transformations it is easy to see that $B^a=i \{Q,\bar{c}^a\}$. 
  
Having recalled these basic facts now we can see how to obtain the Ward Identities given in eq.~\eqref{qcd_id}.
Consider the reduction formula for a $S$ matrix element where $n$ gluon asymptotic states have been removed and replaced by momentum contraction, leaving $m$ incoming and $p$ outgoing physical states.
Taking the Fourier transform (${\cal F}$) one has
\bea
&{}&\!\!\!\!\!\!\!q_1^{\mu_1} \cdots q_n^{\mu_n} 
M_{\mu_1 \mu_2 \cdots \mu_n}(q_1,\cdots , q_n; k_1, \cdots, k_{m+p})=\nonumber\\
&{}&{\cal F} \left[  \Box_{x_1} \cdots  \Box_{x_n} \langle phys_{out} |T  \partial^\mu v_\mu(x_1)\cdots \partial^\mu v_\mu(x_n) | phys_{in}\rangle\right]=
\nonumber\\
&{}&\xi^n {\cal F} \left[  \Box_{x_1} \cdots  \Box_{x_n} \langle phys_{out} | T B (x_1)\cdots B(x_n) | phys_{in}\rangle\right]=\nonumber\\
&{}&(i \xi)^n {\cal F} \left[  \Box_{x_1} \cdots  \Box_{x_n} \langle phys_{out} | T \{Q,\bar{c}
(x_1)\}\cdots \{Q,\bar{c}(x_n)\}) | phys_{in}\rangle\right]=0.
\label{Ward_brst}
\eea
The $\Box_{x_i}$'s are introduced to keep track of the standard relation 
between Green's functions and S matrix elements but they do not play any role in deriving the identity.
Let us stress that the momenta corresponding to the dependence in $x_1, \cdots x_n$ are not on shell.
The expression is zero thanks to the nilpotent property of $Q$ and to the fact that $Q |phys\rangle=0$.
\subsection{Effective action with reggeized gluons}
Let us now consider the case of our interest,  the Effective Action~\cite{effact} which includes reggeized gluons.
This action is non local, and it has been constructed in such a way that it is gauge invariant, 
including the reggeized gluons (which are not on mass shell) as external states. In order to achieve this one has to introduce, order by order in perturbation theory, a well-defined  set of induced interactions of reggeized gluons with normal gluons. All these induced interactions, as well as the conventional QCD interactions,  are contained in the gauge invariant effective action for gluons and reggeized gluons.

Let us write the effective action which describes the coupling of the reggeized gluons to a cluster centered at rapidity $y_0$ where all the produced particles belong to a rapidity interval $\eta \ll \log s$, which means that $|y-y_0| < \eta$. We denote the gluon field as $v_\mu(x)=-i T^a v^a_\mu(x)$ and the reggeized gluons fields as $A_\pm(x)=-i T^a A^a_\pm(x)$, which satisfy the kinematical constraints
\be
\partial_\mp A_\pm=0 \,,
\label{kin_cons}
\ee
according to the quasi-multi-reggeon kinematics.
The reggeized gluons are characterized by $n^{+}_\mu$ and $n^{-}_\mu$ polarization vectors.

The effective action for a given rapidity interval $\eta$ reads~\cite{effact} 
\be
S=\int {\rm d}^d x \left[ {\cal L}(v_\mu) 
 -{\rm Tr} 
\left( V_+ \Box A_- +V_- \Box A_+ \right)+ 2\,{\rm Tr}\left(A_+\Box A_-\right) \right],
\label{eff_act}
\ee 
Here the first term coincides with the gauge fixed QCD lagrangian 
in~\eqref{lag_QCD}, and the second one describes the interaction of the 
reggeized gluons fields with the gluons  through the induced terms:
\be
V_\pm(v) =-\frac{1}{g} \partial_\pm P \exp \left( -\frac{g}{2} \int_{-\infty}^{x^{\pm}} v_\pm(y) d y^\pm\right) =
v_\pm-g v_\pm \frac{1}{\partial_\pm} v_\pm + g^2 v_\pm \frac{1}{\partial_\pm} v_\pm \frac{1}
{\partial_\pm} v_\pm - \, \cdots \,\,,
\label{induced}
\ee
In contrast to these interactions terms which are local in rapidity, the 
last term, the kinetic term for the reggeized gluons, describes the 
interaction of particles with different rapidities. 
We note that, in this action, the reggeized gluons play a role similar to 
classical fields, i.e. within each rapidity cluster they do not appear in loops. For the last two terms in Eq.~\eqref{eff_act} we can also introduce a more compact notation. We define 
\bea
&&A_\mu=\frac{1}{2}\left(A^+ (n^-)_\mu+A^-(n^+)_\mu\right)=2\left(A_+ (n_-)_\mu+A_-(n_+)_\mu\right)\nonumber\\
&&V_\mu=\frac{1}{2}\left(V^+ (n^-)_\mu+V^-(n^+)_\mu\right)=2 \left(V_+ (n_-)_\mu+V_-(n_+)_\mu\right)
\label{fullfields}
\eea
with the normalization $n^+\cdot n^-=2$ and $n_+\cdot n_-=\frac{1}{2}$. Note that we have $\partial^\mu A_\mu=0$, due to Eq.~\eqref{kin_cons}, which is similar to the Lorentz condotion for the real gluons.
This  allows to write, in Eq.~\eqref{eff_act}, the kinetic term of the reggeized gluon as $\frac{1}{2} {\rm Tr} \left( A \Box A \right)$ and the induced interaction part  as
$-\frac{1}{2} {\rm Tr} \left( V \Box A \right)$.

Under gauge transformations one has $\delta v_\mu = \left[ D_\mu , \chi \right]$ which implies on the induced terms a variation
$\delta V_\pm=\partial_\pm \left[ \chi, (D_\pm)^{-1} \right]$. 
Therefore, recalling the kinematical constraints in Eq.~\eqref{kin_cons}, one notes that, after integration by parts, the variation of the terms in the traces
is zero provided the function $\chi(x)$ which describes the gauge transformation vanishes as $x\to \infty$ and $\delta A_\pm=0$.
Apart from the gauge fixing and ghost terms we have therefore a fully gauge invariant effective action by considering gauge invariant reggeized gluons. After introducing gauge fixing and ghost terms this action enjoys, similarly to the normal QCD case, a global BRST symmetry
with associated conserved canonical charge $Q$ such that $\delta_\ep A_\pm=0$.

The objects in which we are interested are the gauge invariant scattering amplitudes which,
in general, involve quarks, on shell physical gluon states and reggeized gluons (the extensions to the $N=4$ SYM case is straightforward).
Such scattering amplitudes are constructed from the Green's functions using the LSZ reduction with respect to the lines with the physical quark
and gluon states.  For example, an amplitude with a physical gluon and two reggeized gluons (the BFKL production vertex) is related to the following Green function
\be
\langle 0 | {\rm T} A_{\mu_1}(x_1) A_{\mu_2}(x_2) v_{\mu_3}(x_3) | 0 \rangle \,.
\label{ex-green}
\ee
Choosing the unphysical polarizations for each of the reggeized gluons and the physical polarization for the normal gluon,
taking the Fourier transform in the LSZ reduction formula and removing for the latter the physical pole, we have the effective BFKL production vertex
\be
M=n^-_{\mu_1}\, n^+_{\mu_2}\,  \ep_{\mu_3}(k) M^{\mu_1\mu_2\mu_3}(q_1, q_2, k) \,,
\label{ex-amp}
\ee
where the polarization vector $n^-$ belongs  to the reggeized gluon with momentum $q_1$ (with a large "$+$" component),
and the vector $n^+$ to the gluon with  momentum $q_2$ (with a large "$-$" component).

Note that this amplitude contains induced terms which couple the reggeized gluons to the usual gluons  
(rhs of ~\eqref{induced}).  These contributions are necessary for restoring  the gauge invariance of the amplitude. 
We shall use two different equivalent notations to write the Ward identities.
In the notation of ~\cite{feynruleffact}, starting from~\eqref{induced}, not all Feynman diagrams are proportional to the product of two external polarization vectors, $n^-_{\mu_1}\, n^+_{\mu_2}$. Induced terms which do not contain the vector 
$n^-$ have to be left out in the Ward identity in $q_1$, terms without $n^+$ do not participate 
in the Ward identity in $q_2$ (such terms come from the second, third,... terms in  Eq.~\eqref{induced}, i.e. from those interactions where the external reggeized gluon couples to 
two or more elementary gluons).
In  Eq.~\eqref{ex-amp} we have introduced a simplified notation which allows to write the full amplitude using a uniform contraction with the $n^\pm$ vectors. This follows from having introduced the general objects of Eq.~\eqref{fullfields} containing both polarizations,
which, after contractions with $n^\pm$, give the specific polarized terms.
 
Let us now formulate, for this example, the Ward identity in the external reggeon line with momentum  $q_1$. 
The argument is analogous to the one used in the previous section for pure QCD and  is based on the gauge invariance of the Effective Action for reggeized gluons. Since we have seen that under a BRST transformation the reggeized gluons do not change,
the corresponding quantum states  $|A_\pm \rangle$ are also annihilated by the BRST operator $Q$ and we can extend the Hilbert space corresponding to gauge invariant states adding also the reggeized gluon states, which can be considered on the mass shell as physical states in the $t$-channel. From it one may construct the multi-particle Fock space. 
As before, therefore, in order to obtain a Ward identity starting from the Green's function in  Eq.~\eqref{ex-green} we consider a new Green's function, obtained replacing at point $x_1$ the reggeized gluon field operator $A_{\mu_1}(x_1)$ by the operator
$\partial^{\mu_1} v_{\mu_1}(x_1)$: 
\be
\langle 0 | {\rm T} \partial^{\mu_1} v_{\mu_1}(x_1)A_{\mu_2}(x_2) v_{\mu_3}(x_3) | 0 \rangle. 
\label{ex-green2}
\ee
Proceeding now in the same way as in Eq.~\eqref{Ward_brst}, using the relation
$\partial^{\mu_1} v_{\mu_1}(x_1) = i \xi \{Q,\bar{c}(x_1)\}$ ,
applying the LSZ reduction to  the physical gluon line and fixing the polarization of the reggeized gluon at point $x_2$, we end up with the following identity
\be
0=q_1^{\mu_1}\, n^+_{\mu_2}\,  \ep_{\mu_3}(k) \tilde M_{\mu_1}^{\mu_2\mu_3} (q_1, q_2, k) \,.
\label{ex-id}
\ee
It is important to note that the replacement $A_{\mu_1}(x_1) \to \partial^{\mu_1} v_{\mu_1}(x_1)$
eliminates all those induced contributions in Eq.~\eqref{induced}, where the reggeized gluon $q_1$ ends on two 
gluons (for example at tree level here one cannot have induced terms of higher order). In the same time the reggeized gluon with momentum $q_2$ interacts with all vertices in Eq.~\eqref{induced}. 

Let us generalize these arguments. Using the same notation as in Eq.~\eqref{ex-amp}, the 
amplitudes are of the form
\be
M=
\ep_1^{\mu_1}(k_1)\cdots \ep_n^{\mu_n}(k_n) n^{-}_{\nu_1} \cdots n^{-}_{\nu_l} n^{+}_{\rho_1} \cdots n^{+}_{\rho_r} 
M_{\mu_1\!\cdots \mu_n}^{\nu_1\!\cdots \nu_l \rho_1\!\cdots \rho_r  }(k_1,\!\cdots\!, k_n, q_1, \!\cdots\!, q_l, q'_1,\!\cdots\!, q'_r), 
\label{amp}
\ee
where the tensor amplitude $M_{\mu_1\cdots \mu_n}^{ \nu_1\cdots \nu_l \rho_1\cdots \rho_r  }(k_1,\cdots, k_n, q_1, \cdots, q_l, q'_1,\cdots, q'_r)$ is constructed from fields of reggeized gluons,
$A_{\nu_1},...,A_{\nu_l}$,  $A_{\rho_1},...,A_{\rho_r}$, and physical gluon fields $v_{\mu_1},...v_{\mu_n}$.
At this stage, the amplitude $M$ contains all the induced interaction associated to the $r$ "$+$" polarized  reggeized gluons and to the $l$ "$-$" polarized  reggeized gluons. 
By suitable replacements $A_\mu(x) \to \partial^\mu v_\mu(x)$ one can now write a tower of Ward-like identities for the reggeized gluons which are analogous to the ones in eq.~\eqref{qcd_id}. They are obtained from Eq.~\eqref{amp} by replacing the corresponding polarization vectors of the reggeized gluons by the contraction with the momentum, taking into account that the tensor amplitude must be substituted by a new one, $\tilde{M}_{\mu_1\cdots \mu_n}^{ \nu_1\cdots \nu_l \rho_1\cdots \rho_r  }$, since the terms corresponding to the induced interactions of the line with contractions must be removed. 
 Indeed, as in our example, the replacement  $A_\mu(x) \to \partial^\mu v_\mu(x)$ eliminates those induced graphs where the reggeized gluon couples to two (or more) gluon fields. In the  notation of ~\cite{feynruleffact}, these are exactly those graphs which do not have a polarization vector $n^{\pm}$ for the reggeized gluon and do not participate in the Ward identity. In a similar way one obtains also Ward identities for the gluon fields.
 
Our Ward identities therefore take the form:
\bea
&{}&0=\ep_1^{\mu_1}(k_1)\cdots k_{i_1}^{\mu_{i_1}}\cdots k_{i_\alpha}^{\mu_{i_\alpha}} \cdots \ep_n^{\mu_n}(k_n) \,
n^{-}_{\nu_1} \cdots q_{m_1}^{\nu_{m_1}}\cdots q_{m_\beta}^{\nu_{m_\beta}} \cdots n^{-}_{\nu_l}\times \nonumber \\
&{}&\!\!\!\!\!\!\!\!\!\!\!\!\!\!n^{+}_{\rho_1} \cdots {q'}_{p_1}^{\rho_{p_1}}\cdots {q'}_{p_\gamma}^{\rho_{p_\gamma}} \cdots  n^{+}_{\rho_r}
\tilde{M}_{\mu_1\cdots \mu_n \nu_1\cdots \nu_l \rho_1\cdots \rho_r  }(k_1,\cdots\!, k_n, q_1, \cdots\!, q_l, q'_1,\cdots\!, q'_r),
\label{EA_id}
\eea
where the subsets $\alpha$, $\beta$ and $\gamma$  of physical polarization states, "-" polarized and "+" polarized reggeized gluons, respectively, have been replaced by the corresponding momentum contractions. We stress that $\tilde{M}$ is related to the modified product of field operators where some of the reggeized gluon fields $A_{\mu}$ have been replaced by gluon fields $v_{\mu}$. 

In the following we will show that these Ward identities can be used to obtain a new (and useful) representation for amplitudes and Green's functions, in which the unphysical polarizations of reggeized gluons are substituted by transverse momentum vectors.

\section{An application: the BFKL kernel}
As a first application, we investigate the use of Ward identities for the real part of the BFKL kernel. 
As a result, we will present a new derivation of the kernel. We proceed as follows. Making use of the effective 
action we begin with the LO 4-point function of four reggeized gluons where, in contrast to the usual derivation 
based on the $s$-channel unitarity, the produced $s$-channel gluon is off shell, and its 
longitudinal momenta (Sudakov or light cone variables) are not yet integrated. At this stage, the reggeized gluons carry unphysical polarizations. 
Beginning with one of the reggeized gluons, we derive a Ward identity and use it to substitute the unphysical 
polarization vector by the corresponding transverse momentum vector. Repeating this procedure, step by step, for the remaining gluons we arrive at a representation of the BFKL kernel in which for all four reggeized gluons the unphysical polarization vectors $n^+$ or $n^-$ are replaced  by transverse momentum vectors. We then show that after the integration over the longitudinal momentum of the $s$-channel gluon this new form of the BFKL kernel coincides with the standard expression. Finally, we re-formulate our results and 
show that they are in agreement with the Ward identities stated in section 1. 
\subsection{Derivation from the effective action}       
We begin with the notation for the effective diagrams shown in Fig.1:
\begin{center}
\epsfig{file=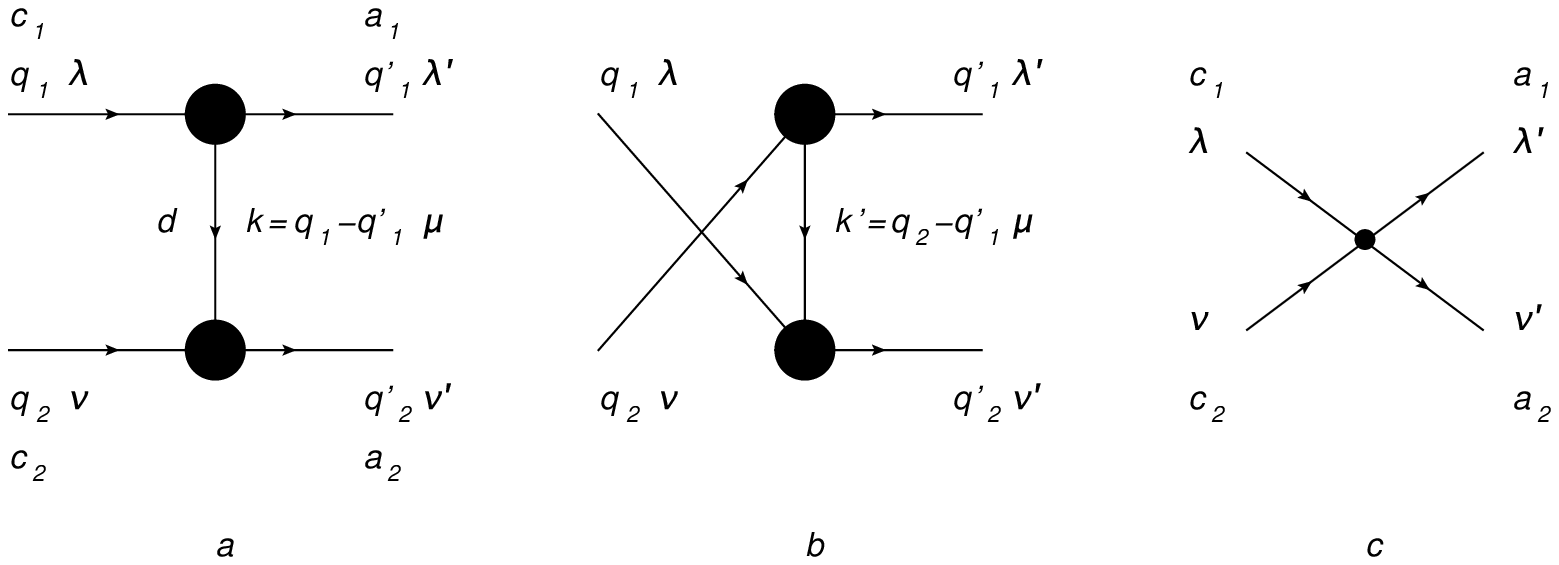,width=14cm,height=4cm}\\
Fig.1: the BFKL kernel.
\end{center}
The 4-point function (which, after integration over the longitudinal component of the momentum $k$, will 
become the BFKL kernel) is obtained from squares of two effective production vertices plus the quartic interaction. 
In contrast to the usual derivation of the BFKL kernel, we take the produced gluon to be off-shell and do the longitudinal integration later. 
The upper effective production vertex in Fig.1a has the form (disregarding, for the moment, 
the color structure):
\bea
C^{\mu}(q_1,q'_1)&=& 2 \left[
\left( \frac{q^+_{1}}{2} -\frac{q_1^2}{{q'_1}^-}\right) (n^-)^{\mu} +
\left( \frac{{q'_1}^-}{2} - \frac{q'^{2}_1}{q_1^+}\right)(n^+)^{\mu}-
 \left(q_1+q'_1\right)_{\perp}^{\mu}
\right].
\eea
Within the effective action the production vertex is derived from the triple gluon Yang-Mills vertex $\gamma$ and from the two induced vertices:
\bea
C^{\mu}(q_1,q'_1)&=& (n^-)_{\lambda} \gamma^{\lambda \mu \lambda'}(q_1,q'_1) (n^+)_{\lambda'}
\nonumber \\ &&-  \left( \frac{q_1^2}{{q'_1}^-} (n^-)^{\mu} (n^-)^{\lambda'} \right)(n^+)_{\lambda'}  -  
    (n^-)_{\lambda}  \left( (n^+)^{\lambda}  (n^+)^{\mu} \frac{q'^{2}_1}{q_1^+}\right)
\label{effvertex}
\eea  
with 
\bea
\gamma^{\lambda \mu \lambda'}(q_1,q'_1)= g^{\mu \lambda}  \left( -\frac{(n^+)^{\lambda'}}{2} {q'_1}^-+
\frac{(n^-)^{\lambda'}}{2} 2q_1^+ +(2 q_1 -q'_1) _{\perp}^{\lambda'}\right) \nonumber \\
+g^{\mu \lambda'}  \left( \frac{(n^+)^{\lambda}}{2} 2{q'_1}^--
\frac{(n^-)^{\lambda}}{2} q_1^+ +(2q'_1 -q_1)_{\perp}^{\lambda}\right) \nonumber \\ 
- g ^{\lambda \lambda'} (q_1+q'_1)^{\mu}.
\eea  
It will be convenient to introduce a short-hand notation for the induced terms\footnote{The subscripts 'L' and 'R' follow the notation of \cite{feynruleffact}: 'left central' and 'right central'}:
\bea
I_R^{\mu \lambda'}(q_1^2, {q'_1}^-)& =& -  \left( \frac{q_1^2}{{q'_1}^-} (n^-)^{\mu} (n^-)^{\lambda'} \right)
\nonumber \\
I_L^{\lambda \mu}({q'_1}^2,q_1^+) &=& -  \left( (n^+)^{\lambda}  (n^+)^{\mu} 
\frac{q'^{2}_1}{q_1^+}\right),
\eea
such that (\ref{effvertex}) takes the form:
\be
C^{\mu}(q_1,q'_1) = (n^-)_{\lambda} \gamma^{\lambda \mu \lambda'}(q_1,q'_1) (n^+)_{\lambda'} +
I_R^{\mu \lambda'} (q_1^2, {q'_1}^-)(n^+)_{\lambda'} + (n^-)_{\lambda} I_L^{\lambda \mu}({q'_1}^2,q_1^+).
\label{prodvertex}
\ee
Contraction with the momentum of the produced gluon leads to:
\bea
(n^-)_{\lambda} \gamma^{\lambda \mu \lambda'}(q_1,q'_1) (n^+)_{\lambda'}  k_{\mu}=
 2 (q_1^2 - {q'_1}^2).
\label{Wardknn}
\eea
Including the induced terms, $I_L$ and $I_R$,  gauge invariance of the production vertex is restored:
\be
C^{\mu}(q_1,q'_1) k_{\mu} = 0.
\label{WardkC}
\ee
Including the color structure, Figs.1a and b  have the color coefficients
\be
f_{c_1da_1} f _{c_2a_2d}
\ee
and
\be
f_{c_2da_1} f _{c_1a_2d}, 
\ee  
resp.
The quartic interaction in Fig.1c provides the same two color structures with the following 
coefficients:
\be
(n^-)_{\lambda} (n^-)_{ \nu} \left( g^{\lambda \nu} g^{\lambda'\nu'}  -  
g^{\lambda' \nu}  g^{\lambda \nu'}   \right)   (n^+)_{\lambda'} (n^+)_{\nu'} = - 4
\label{quartic1}     
\ee
and
\be
(n^-)_{\lambda} (n^-)_{ \nu} \left(g^{\lambda \nu} 
g^{\lambda'  \nu'} - g^{\lambda \lambda'}  g^{\nu \nu'}  \right)  (n^+)_{\lambda'} (n^+)_{\nu'} = - 4.
\label{quartic2}
\ee
The third color structure of the quartic coupling, $f_{c_1c_2l} f_{la_2a_1}$,  will be disregarded 
since it correspond to the color octet $t$-channel.  Putting everything together we find, from 
Fig.1a,
\be
\frac{1}{k^2} C^{\mu}(q_1,q'_1) C_{\mu}(q_2,q'_2) + quartic\,\,\, coupling =
\frac{8}{k^2} \left( q^2 + \frac{q_1^2 {q'_2}^2 + q_2^2 {q'_1}^2}{k^+ k^-}\right)
\label{BFKL1}
\ee
(note that, on the rhs of this equation, $k^2= k_{\mu} k^{\mu}= k^+ k^- +k_{\perp}^2$ denotes the square of a four vector;
all other  squared momenta are effectively purely transverse). Without the piece of the quartic coupling  
there would be the additional term  $\frac{k^2}{2}$ inside the bracket. 
We write (\ref{BFKL1}) also in another form:
\bea
(2.29)& =& 8\left(\frac{q^2}{k^+k^-} + \frac{1}{k^2} \frac{- k_{\perp}^2 q^2 + 
(q_1^2 {q'_2}^2 + q_2^2 {q'_1}^2)}{k^+ k^-}\right).
\label{BFKL1prime1}
\eea
Here the numerator of the second term has the property that it vanishes 
whenever any of the transverse momenta $q_1$, $q_2$, $q'_1$, or $q'_2$ goes to zero.
In order to make this 'zero property' explicit, we list a few alternative 
forms of the vertex:
\bea
(2.29)& =& 8 \left(\frac{q^2}{k^+k^-} + \frac{2}{k^2} 
\frac{(q_{1\perp}q_{2\perp}) \; (q'_{1\perp} q'_{2\perp})  +(q_{1\perp}\times q_{2\perp})\;  (q'_{1\perp}\times q'_{2\perp})}{k^+ k^-}\right)
\nonumber\\
& =& 8 \left(\frac{q^2}{k^+k^-} + \frac{2}{k^2} 
\frac{(q_{1\perp}q'_{1\perp}) \; (q_{2\perp}q'_{2\perp})  +(q_{1\perp}\times q'_{1\perp})\;  (q_{2\perp}\times q'_{2\perp})}{k^+ k^-}\right)
\nonumber\\
&=& 8 \left(\frac{q^2}{k^+k^-} + \frac{1}{k^2} \frac{q_1 q_2^* {q'_1}^* q'_2 
+ q_1^* q_2 q'_1 {q'_2}^*
}{k^+ k^-}\right)\,.
\label{BFKL1prime2}
\eea
In the first two lines we have introduced a cross product-like 
notation $(q_{1\perp} \times q_{2\perp})=q_{1x}q_{2y}-q_{1y}q_{2x}$, while
in the last line we have used complex notation
where $p=p_x+i p_y$ (note that, following the Sudakov notation, we use
$p_\perp^\mu =(0,0,p_x,p_y)$). 
Because of $k^+ = - {k'}^+$, $k^- = {k'}^-$ 
one easily recognizes the cancellation of the first term, once we add the crossed graph. 
We shall show in sections $2$ and $3$ that, after the use of the Ward identities, 
the first piece, $8\frac{q^2}{k^+k^-}$, which destroys the zero property will be absent even before adding the crossed graph.

Before we study the integration over the longitudinal variable, $k^-=-q'_{1}{}^-$,
we mention another important feature of the representations (\ref{BFKL1prime2}).
In this form, the BFKL kernel is closely related to the Weizs\"acker-Williams approximation in  which the interaction of the reggeized gluons, at vanishing transverse momenta, can be expressed in terms of on-shell gluon scattering amplitudes. 
   
Next we turn to the integration over the longitudinal variable, $k^-=-q'_{1}{}^-$. Let us fix, in (\ref{BFKL1prime1}), 
$k^+$ at some positive value. As we said before, the first term, $8\frac{q^2}{k^+k^-}$, 
cancels after adding the contribution from the crossed graph.
The second term with the additional denominator $1/k^+k^-$ converges for large $k^-$, 
but the singularity at $k^-=0$ needs to be regularized. From the analysis of Feynman graphs we 
find an infrared cutoff $|k^-|>\mu^2/\sqrt{s}$ which implies the following definition of the $k^-$ integral:
\be
I= \int \frac{dk^-}{k^+k^-}\frac{1}{k^+k^-+k^2_{\perp}+i\epsilon} := 
\left( \int_{-\infty}^{-\mu^2/\sqrt{s}} + \int_{\mu^2/\sqrt{s}}^{\infty} \right)  
\frac{dk^-}{k^+k^-}\frac{1}{k^+k^- +k^2_{\perp}+i\epsilon}, 
\label{intreg1}
\ee
where $\mu^2$ denotes a momentum scale of the order of the transverse momenta. 
Adding and subtracting the contribution of a small semicircle in the upper half complex plane, in the first case
the upper half plane has no singularity, and the integral vanishes. 
From the subtraction of the semicircle we are left with     
\be 
I= \frac{i\pi}{k^+ k^2_{\perp}}.
\label{intreg2}
\ee
The same result is obtained if we add and subtract a semicircle in the lower half plane. 
Obviously, this result is independent of any $i\epsilon$ description of the pole at $k^- =0$. 
Proceeding in the same way for the region $k^+<0$, we find 
\be 
I= - \frac{i\pi}{k^+ k^2_{\perp}}.
\label{intreg3}
\ee
Including the integration over $k^+$ we find, for the sum over both regions,
\be
\int dk^+ I = \frac{2i\pi}{k^2_{\perp}} \int_{\mu^2/\sqrt{s}}^{\sqrt{s}} \frac{dk^+}{k^+}.
\ee   

The same result is obtained if we use the principal value prescription which follows from the derivation of the effective BFKL production vertex  
from Feynman graphs: the two graphs which describe the radiation of the 
produced $s$-channel gluon from the vertices to the left of the BFKL vertex imply: 
\be
\int \frac{dk^-}{k^-} \to  \int dk^- \frac{1}{2} \left(\frac{1}{k^-+i\epsilon} +  \frac{1}{k^- -i\epsilon} \right).
\label{P-value}
\ee     
When applied  to the second part of (\ref{intreg1}), we again arrive at (\ref{intreg2}) 
\be
\int \frac{dk^-}{k^2+i \epsilon} \frac{q_1^2 {q'_2}^2 + q_2^2 {q'_1}^2}{k^+ k^-} = \frac{i\pi}{k^+}  \frac{q_1^2 {q'_2}^2 + q_2^2 {q'_1}^2}{k^2_{\perp}}.
\ee  
Note that both our symmetric choice of the infrared cutoff in (\ref{intreg1}) and 
the principal value prescription in (\ref{P-value}) are related the fact that we are considering even signature amplitudes.
 
For the crossed graph we interchange $q_1$ and $q_2$ and replace $k=q_1-{q'}_1$ by $k'=q_2-{q'}_1$. Using ${k'}^+= -k^+$ we obtain:
\be
\int \frac{d{k'}^-}{{k'}^2+i \epsilon} \frac{q_1^2 {q'_1}^2 + q_2^2 {q'_2}^2}{{k'}^+ {k'}^-} = 
\frac{i\pi}{{k'}^+}  \frac{q_1^2 {q'_1}^2 + q_2^2 {q'_2}^2}{{k'}^2_{\perp}}.
\label{intreg4}
\ee  
In this way, the BFKL kernel is crossing symmetric under the exchange $q_1 \to q_2$.

We conclude by mentioning that we would have obtained the same result for the longitudinal integrations by using the identity 
\be
\frac{1}{k^2 + i \epsilon}= {\cal P} \frac{1}{k^2} - i\pi \delta(k^2).
\ee    
The principal value term vanishes since we integrate over positive and negative 
values of $k^+$ and $k^-$.
\subsection{Ward identities on the rhs, $q'_1$ and $q'_2$}
In the following we search for an alternative expression which improves the convergence in 
the longitudinal component,  $q'_{1}{}^-$, 
To this end we replace the $(n^+)$ and $(n^-)$ vectors of the $t$-channel gluons by transverse momenta. We begin with the two gluons on the rhs, and in second step, apply the same procedure 
to the gluons on the lhs.  We begin with the Ward identity in the upper $t$-channel gluon on the rhs
with momentum $q'_1$. We find:
\bea 
\left((n^-)_{\lambda} \gamma^{\lambda \mu \lambda'}(q_1,q'_1) + I_R^{\mu \lambda'}(q_1^2,{q'_1}^-) \right)
{q'_1}_{ \lambda'} = - k^2 (n^-)^{\mu} -{q'_1}^- k^{\mu}
\label{Wardq1p(ng+IL)}
\eea
(if the produced gluon were on mass shell, and we would multiply with a physical polarization vector the rhs would vanish).
When contracting the rhs, in Fig.1a, with the lower production vertex, the piece proportional to 
$k^{\mu}$ vanishes because of the gauge invariance property of the lower vertex.
We thus are left with:
\be
\frac{1}{k^2} C_{\mu}(q_2,q'_2) \left((n^-)_{\lambda} \gamma^{\lambda \mu \lambda'} (q_1,q'_1)+ I_R^{\mu \lambda'}(q_1^2,{q'_1}^-) \right) {q'_1}_{ \lambda'}  = -2(q'_2)^- + 4 \frac{{q'_2}^2}{q_2^+}.
\label{Ward1}
\ee  
The first term cancels if we add the contribution from the quartic coupling, obtained from
(\ref{quartic1}) by replacing $(n^+)_{\lambda'}$ by  ${q'_1}_{\lambda'}$ and observing 
${q'_1}^- =- {q'_2}^-$.
Similarly, the crossed graph in Fig.1b yields:
\bea 
\left((n^-)_{\lambda} \gamma^{\lambda \mu \lambda'}(q_2,q'_1) + I_R^{\mu \lambda'}(q_2^2,{q'_1}^-) \right)
{q'_1}_{ \lambda'} = - {k'}^2 (n^-)^{\mu} -{q'_1}^- {k'}^{\mu}
\eea
and 
\be
\frac{1}{{k'}^2}C_{\mu} (q_1,q'_2)\left((n^-)_{\lambda} \gamma^{\lambda \mu \lambda'}(q_2,q'_1) + I_R^{\mu \lambda'}(q_2^2,{q'_1}^-) \right) {q'_1}_{ \lambda'}  = -2  (q'_2)^- + 4 \frac{{q'_2}^2}{q_1^+}.
\label{Ward2}
\ee  
Again, the first term vanishes if we add the quartic coupling. The second pieces of (\ref{Ward1})
(\ref{Ward2}) sum up to zero, since $q_1^+ +q_2^+\simeq 0$.

As a result, we have verified the following Ward identity:
\begin{align}
\Big[ \frac{1}{k^2}  \left((n^-)_{\lambda} \gamma^{\lambda \mu \lambda'}(q_1,q'_1) + I_R^{\mu \lambda'}(q_1^2,{q'_1}^-) \right) C_{\mu}(q_2,q'_2) + \nonumber\\
\frac{1}{{k'}^2} \left((n^-)_{\lambda} \gamma^{\lambda \mu \lambda'}(q_2,q'_1) + I_R^{\mu \lambda'} (q_2^2,{q'_1}^-)\right) C_{\mu}(q_1,q'_2) +\,\,
quartic \,\,couplings \Big] {q'_1}_{ \lambda'}  =0.
\end{align}
Therefore, in 
\begin{align}
\Big[ \frac{1}{k^2}  \left((n^-)_{\lambda} \gamma^{\lambda \mu \lambda'}(q_1,q'_1) + I_R^{\mu \lambda'}(q_1^2,{q'_1}^-) \right) C_{\mu}(q_2,q'_2) + \nonumber\\
\frac{1}{{k'}^2} \left((n^-)_{\lambda} \gamma^{\lambda \mu \lambda'}(q_2,q'_1) + I_R^{\mu \lambda'}(q_2^2,{q'_1}^-) \right) C_{\mu}(q_1,q'_2) +\,\,
quartic \,\,couplings \Big] (n^+)_{ \lambda'}  
\end{align}
we  can substitute ($q'_1 = \frac{n^+}{2} {q'_1}^- + q'_{1 \perp}$) 
\be
(n^+)_{\lambda' } \rightarrow 
-2 \frac{(q'_{1\perp})_{ \lambda'}  }{{q'_1}^-}.
\label{replace}
\ee
Taking into account that a purely transverse vector, contracted with $I_L$ or with the quartic coupling 
gives zero contribution, the 4-point function can written in the following form: 
\bea
\Big[ - \frac{2}{{q'_1}^-} (n^-)_{\lambda} \gamma^{\lambda \mu \lambda'}(q_1,q'_1) (q'_1)_{\perp \lambda'} + (n^-)_{\lambda} I_L^{\lambda  \mu}({q'_1}^2,q_1^+) \Big]
C_{\mu}(q_2,q'_2) \nonumber \\ 
 +\Big[- \frac{2}{{q'_1}^-} (n^-)_{\lambda} \gamma^{\lambda \mu \lambda'}(q_2,q'_1) (q'_1)_{\perp \lambda'} + (n^-)_{\lambda} I_L^{\lambda  \mu}({q'_1}^2,q_2^+)
\Big]
C_{\mu}(q_1,q'_2).
\label{BFKL1.5}
\eea
Here 
\be
(n^-)_{\lambda} \gamma^{\lambda \mu \lambda'}(q_1,q'_1) (q'_1)_{\perp \lambda'} = (n^-)^{\mu} (2q_1-q'_1)_{\perp}\cdot q'_{1\perp} + 2{q'_1}^- (q'_{1 \perp})^{\mu},
\label{ngt}
\ee
and for the produced gluon we have the Ward identity
\be
\Big[ - \frac{2}{{q'_1}^-} (n^-)_{\lambda} \gamma^{\lambda \mu \lambda'}(q_1,q'_1) (q'_1)_{\perp \lambda'} + (n^-)_{\lambda} I_L^{\lambda  \mu}({q'_1}^2,q_1^+) \Big] k_{\mu} = 0.
\label{Wardk(ngt+IR)}
\ee
apart from the usual Ward identity for $C^\mu$ (see Eq.~\eqref{WardkC}).
Next we proceed to the Ward identity in the lower gluon, $q'_2$. In analogy with (\ref{Wardq1p(ng+IL)}) we 
have
\bea 
\Big[(n^-)_{\nu} \gamma^{\nu \mu \nu'}(q_2,q'_2) + I_R^{\mu \nu'}(q_2^2,{q'_2}^-) \Big]
{q'_2}_{ \nu'} = - k^2 (n^-)^{\mu} +{q'_2}^- k^{\mu}.
\label{Wardq2p(ng+IL)}
\eea
After contraction with the square bracket in the first line of (\ref{BFKL1.5}) we obtain
\bea
\Big[ - \frac{2}{{q'_1}^-} (n^-)_{\lambda} \gamma^{\lambda \mu \lambda'}(q_1,q'_1) (q'_1)_{\perp \lambda'} + (n^-)_{\lambda} I_L^{\lambda  \mu}({q'_1}^2,q_1^+) \Big]  
\Big[(n^-)^{\nu} \gamma_{\nu \mu \nu'}(q_2,q'_2) + I_{R\,\, \mu \nu'} (q_2^2,{q'_2}^-)\Big]
{q'_2}{ \nu'}  \nonumber
\eea
\be
= 4 k^2 \frac{{q'_1}^2}{q_1^+}.
\ee
A similar treatment of the crossed graph leads to   
\bea
\Big[ - \frac{2}{{q'_1}^-} (n^-)_{\lambda} \gamma^{\lambda \mu \lambda'} (q_2,q'_1)(q'_1)_{\perp \lambda'} + (n^-)_{\lambda} I_L^{\lambda  \mu}({q'_1}^2,q_2^+) \Big]  
\Big[(n^-)^{\nu} \gamma_{\nu \mu \nu'}(q_1,q'_2) + I_{R\,\,\mu \nu'}(q_1^2,{q'_2}^-) \Big]
{q'_2}^{ \nu'}  \nonumber
\eea
\be
= 4 k^2 \frac{{q'_1}^2}{q_2^+},
\ee
and the sum of both cancels. Repeating the steps as described above in Eqs~\eqref{replace}-\eqref{ngt} we find for the 
4-point function:
\bea
\frac{1}{k^2} \Big[ - \frac{2}{{q'_1}^-} (n^-)_{\lambda} \gamma^{\lambda \mu \lambda'}(q_1,q'_1) (q'_1)_{\perp \lambda'} + (n^-)_{\lambda} I_L^{\lambda  \mu}({q'_1}^2,q_1^+) \Big] \hspace{3cm}
\nonumber \\
\cdot \Big[ - \frac{2}{{q'_2}^-} (n^-)^{\lambda} \gamma_{\lambda \mu \lambda'} (q_2,q'_2)(q'_2)_{\perp}^{ \lambda'} + (n^-)^{\lambda} I_{L \,\,\lambda  \mu}({q'_2}^2,q_2^+) \Big] 
\nonumber\\
+ \frac{1}{{k'}^2}\Big[- \frac{2}{{q'_1}^-} (n^-)_{\lambda} \gamma^{\lambda \mu \lambda'} (q_2,q'_1)(q'_1)_{\perp \lambda'} + (n^-)_{\lambda} I_L^{\lambda  \mu}({q'_1}^2,q_2^+)\Big] \hspace{3cm}
 \nonumber \\
\cdot
\Big[- \frac{2}{{q'_2}^-} (n^-)^{\lambda} \gamma_{\lambda \mu \lambda'}(q_1,q'_2) (q'_2)_{\perp}^{ \lambda'} + (n^-)^{\lambda} I_{L\,\,\lambda  \mu}({q'_1}^2,q_1^+) \Big].
\label{BFKL2}
\eea 
Note that in this expression the quartic contribution no longer appears, and the divergence 
$\sim k^2$, in (\ref{BFKL1}) is no longer present.

It is interesting to compare (\ref{BFKL2}) with the known result. After some algebra we find for the  first product on the rhs:
\bea
\frac {1}{k^2} \Big[8q^2 - 8({q'_1}^2 + {q'_2}^2) \left(1+\frac{k_{\perp}^2}{k^+k^-}\right) + 8 \frac{q_1^2 {q'_2}^2 + {q'_1}^2 q_2^2}{k^+k^-}\Big]. 
\label{interm_half_kernel}
\eea 
The integral over ${q'}_1^-$, however,  is still logarithmically divergent, and only after adding the  analogous expression for the crossed graph this divergence cancels and we are allowed to close the integration contour. 

The new representation Eq.\eqref{interm_half_kernel} of the 4-point function differs from our starting expression Eq.~\eqref{BFKL1} only through the second term which can be written as
 $-8 ({q'_1}^2 + {q'_2}^2) /k^+ k^-$. One can easily check that, when adding the 
 crossed  term (obtained by interchanging $q_1$ and $q_2$), we find 
 \be
 -8 ({q'_1}^2 + {q'_2}^2) /k^+ k^- -8 ({q'_1}^2 + {q'_2}^2) /{k'}^+ {k'}^- = 0,
 \ee
 since ${k'}^+ = -k^+$ and ${k'}^- = k^-$. 
As a result,  the sum of Eq.\eqref{interm_half_kernel}) plus its crossed graph equals the sum of Eq.~\eqref{BFKL1} plus its crossed counterpart.
\subsection{Ward identities on the lhs, $q_1$ and $q_2$}
In the final step we start from (\ref{BFKL2}) and make use of Ward identities in the two gluons on the left hand.  Beginning with $q_1$ we first compute 
\bea
(q_1)_{\lambda}  \gamma^{\lambda \mu \lambda'}(q_1,q'_1) (q'_1)_{\perp \lambda'} 
\!\!\!&=& k^{\mu} q_{1\perp}\cdot q'_{1\perp} -
k^2 {q'_{1\perp}}^{\mu}  +{q'_1}^2 ({q'_{1\perp}}^{\mu} - q_1^{\mu}) \nonumber \\
&= &k^{\mu}\left(  q_{1\perp}\cdot q'_{1\perp} - {q'_1}^2 \right)
-k^2 {q'_{1\perp}}^{\mu}  - \frac{(n^+)^{\mu}}{2} {q'_1}^2 {q'_1}^-.
\eea 
Together with the induced term we get: 
\bea
(q_1)_{\lambda}  \Big[ -\frac{2}{{q'_1}^-} \gamma^{\lambda \mu \lambda'}(q_1,q'_1) (q'_1)_{\perp \lambda'} +  I_L^{\lambda  \mu} ({q'_1}^2,q_1^+)\Big] =-\frac{2}{{q'_1}^-} \left(k^{\mu}\left( q_{1\perp}\cdot q'_{1\perp}- {q'_1}^2\right) - k^2 {q'_{1\perp}}^{\mu} \right).
\label{Ward3}
\eea
and 
\bea
(q_1)_{\lambda}  \Big[ -\frac{2}{{q'_1}^-} \gamma^{\lambda \mu \lambda'}(q_1,q'_1) (q'_1)_{\perp \lambda'} +  I_L^{\lambda  \mu}({q'_1}^2,q_1^+) \Big]\nonumber\\
\Big[ - \frac{2}{{q'_2}^-} (n^-)^{\lambda} \gamma_{\lambda \mu \lambda'}(q_2,q'_2) (q'_2)_{\perp}^{ \lambda'} + (n^-)^{\lambda} I_{L\,\,\lambda  \mu}({q'_2}^2,q_2^+) \Big] 
=-8  k^2 \frac{q'_{1\perp}\cdot q'_{2\perp}}{{q'_1}^-}.
\eea
The crossed graph (second term in (\ref{BFKL2})) gives the same result with the denominator 
 replaced by ${q'_2}^-$, i.e. in the sum both contributions cancel. The replacement 
 (\ref{replace}) leads to 
 \bea
\frac{1}{k^2} \Big[ \frac{4}{q_1^+ {q'_1}^-}   (q_1)_{\perp \lambda} \gamma^{\lambda \mu \lambda'} (q_1,q'_1)(q'_1)_{\perp \lambda'}  \Big] \hspace{3cm}
\nonumber \\
\cdot \Big[ - \frac{2}{{q'_2}^-} (n^-)^{\lambda} \gamma_{\lambda \mu \lambda'}(q_2,q'_2) (q'_2)_{\perp}^{ \lambda'} + (n^-)^{\lambda} I_{L\,\,\lambda  \mu}({q'_2}^2,q_2^+) \Big] 
\nonumber\\
+ \frac{1}{{k'}^2}\Big[- \frac{2}{{q'_1}^-} (n^-)_{\lambda} \gamma^{\lambda \mu \lambda'} (q_2,q'_1)(q'_1)_{\perp \lambda'} + (n^-)_{\lambda} I_L^{\lambda  \mu}({q'_1}^2,q_2^+)\Big] \hspace{3cm}
 \nonumber \\
\cdot
\Big[\frac{4}{{q_1^+q'_2}^-} (q_1)_{ \perp}^{\lambda} \gamma_{\lambda \mu \lambda'}(q_1,q'_2) (q'_2)_{\perp}^{ \lambda'} \Big]
\label{BFKL2a}
\eea
with 
\bea
 \left(   (q_1)_{\perp \lambda} \gamma^{\lambda \mu \lambda'}(q_1,q'_1) (q'_1)_{\perp \lambda'} \right) \nonumber\\ = (q_{1\perp})^{\mu} (2q_1-q'_1)_{\perp} q'_{1\perp} + (q'_{1\perp})^{\mu} (2q'_1-q_1)_{\perp} q_{1\perp} - (q_1 +q'_1)^{\mu} q_{1\perp}q'_{1\perp}
\label{tgt}
\eea
and the Ward identity
\be
k_{\mu} \left(   (q_{1 \perp})_{\lambda} \gamma^{\lambda \mu \lambda'}(q_1,q'_1) (q'_{1 \perp})_ {\lambda'} \right) = 0.
\label{Wardktgt}
\ee
Finally the gluon $q_2$. Starting from the first product of (\ref{BFKL2a}) we need to calculate:
 \bea
\frac{1}{k^2} \Big[ \frac{4}{q_1^+ {q'_1}^-}   (q_{1\perp})_{ \lambda} \gamma^{\lambda \mu \lambda'} (q_1,q'_1)(q'_{1\perp})_{ \lambda'}  \Big] \nonumber \\
\cdot \Big[ - \frac{2}{{q'_2}^-} (q_2)^{\lambda} \gamma_{\lambda \mu \lambda'}(q_2,q'_2) (q'_{2\perp})^{ \lambda'} + (q_2)^{\lambda} I_{L\,\,\lambda  \mu}({q'_2}^2,q_2^+) \Big]. 
\eea
Using (\ref{Ward3}) we obtain, after some algebra:
\bea
\frac{8 }{q_1^+ {q'_1}^- {q'_2}^-}   (q_{1\perp})_{\lambda} \gamma^{\lambda \mu \lambda'}(q_1,q'_1) (q'_{1\perp})_{ \lambda'}  (q'_2)_{\perp \mu}   = \frac{8 }{q_1^+ {q'_1}^- {q'_2}^-} 
\left(q_{1\perp}  q_{2\perp} \,\,   q'_{1\perp}   q'_{2\perp}
- q_{1\perp}  q'_{2\perp} \,\,   q_{2\perp}   q'_{1\perp}\right).
\label{Ward5}
\eea
Similarly, from the second product of (\ref{BFKL2a}) we find:
\bea
\frac{8 }{q_1^+ {q'_1}^- {q'_2}^-}   (q_{1\perp})_{ \lambda} \gamma^{\lambda \mu \lambda'}(q_1,q'_2) (q'_{1\perp})_{ \lambda'}  (q'_{1\perp})_{ \mu}   = \frac{8 }{q_1^+ {q'_1}^- {q'_2}^-} 
\left(q_{1\perp}  q_{2\perp} \,\,   q'_{1\perp}   q'_{2\perp}
- q_{1\perp}  q'_{1\perp} \,\,   q_{2\perp}   q'_{2\perp}\right).
\label{Ward6}
\eea
The sum of the rhs of (\ref{Ward5}) and (\ref{Ward6}) is easily recognized as the sum of the quartic 
couplings (\ref{quartic1}) and (\ref{quartic2}) :
\bea
\frac{8 }{q_1^+ {q'_1}^- {q'_2}^-}  (q_{1\perp}) _{\lambda} (q_{2\perp}) _{\nu}
\Big[ \left( g^{\lambda \nu} g^{\lambda'\nu'}\!-\!g^{\lambda' \nu}  g^{\lambda \nu'}   \right)\!+\! \left(g^{\lambda \nu} 
g^{\lambda'  \nu'}\!-\!g^{\lambda \lambda'}  g^{\nu \nu'}  \right) \Big] 
(q'_{1\perp}) _{\lambda'} (q'_{2\perp}) _{\nu'}.
\eea
This means that the Ward identity in the fourth leg requires an in homogenous term, namely the 
quartic coupling multiplied by the four transverse momenta. 
With this term being included, our final expression for the 4-point function of 4 reggeized gluons becomes:
\bea
 \frac{16 }{q_1^+ q_2^+ {q'_1}^- {q'_2}^-}  (q_{1\perp}) _{\lambda} (q_{2\perp}) _{\nu} 
 \Big[ \frac{1}{k^2} \gamma^{\lambda \mu \lambda'} (q_1,q'_1) g_{\mu \rho} \gamma^{\nu \rho \nu'} (q_2,q'_2)+ \frac{1}{{k'}^2}\gamma^{\nu \mu \lambda'}(q_2,q'_1) g_{\mu \rho}\gamma^{\lambda \rho \nu'}(q_1,q'_2)\nonumber \\
+ \left( g^{\lambda \nu} g^{\lambda'\nu'}\!-\!g^{\lambda' \nu}  g^{\lambda \nu'}   \right)\!+\! \left(g^{\lambda \nu} 
g^{\lambda'  \nu'}\!-\!g^{\lambda \lambda'}  g^{\nu \nu'}  \right)
 \Big]
(q'_{1\perp}) _{\lambda'} (q'_{2\perp}) _{\nu'}. 
 \hspace{2cm}
 \label{BFKL3}
 \eea 
 
The re-appearance of the quartic coupling can be understood from the observation that, in the 
high energy limit,  quartic couplings give nonzero contributions only in a few special cases: for 
example,  when multiplied with  two  $(n^-)$ and two $(n^+)$ vectors  (as it is done in (\ref{quartic1}) and (\ref{quartic2})) 
or when multiplied with four transverse momenta (as it is done in (\ref{BFKL3}). 
Other  'mixed' cases (as in (\ref{BFKL2}): two $(n^-)$ vectors from the left, two transverse momenta form the right) 
lead to vanishing contributions. 
Inspecting \eqref{BFKL3} one can see that the "zero property" mentioned before is manifestly realized separately for each contribution of a single permutation of the external lines.

Finally we present the explicit expression for (\ref{BFKL3}). After some 
algebra one finds for the first part:
 \bea
 (q_{1\perp}) _{\lambda} (q_{2\perp}) _{\nu} 
 \gamma^{\lambda \mu \lambda'}(q_1,q'_1)  g_{\mu \rho} \gamma^{\nu \rho \nu'}(q_2,q'_2) (q'_{1\perp}) _{\lambda'} (q'_{2\perp}) _{\nu'}=\nonumber \\
 (k^+k^- + k_{\perp}^2) q_{1\perp} q'_{1\perp}\,\, q_{2\perp}  q'_{2\perp}  + \frac{1}{2} \left( ((k_{\perp}^2)^2 q^2 - k_{\perp}^2 (q_1^2 {q'_2}^2 +
 {q'_1}^2 q_2^2) \right).
 \eea 
Together with the contribution of the quartic coupling we have:
\bea
\frac{8}{(k^+ k^-)^2} \Big[ \frac{k_{\perp}^2}{k^2} \left( k_{\perp}^2 q^2 - (q_1^2 {q'_2}^2 +
 {q'_1}^2 q_2^2) \right) 
-  \left( k_{\perp}^2 q^2 - (q_1^2 {q'_2}^2 +
 {q'_1}^2 q_2^2) \right) \Big], 
 \label{BFKL3a}
 \eea
which can be simplified into:
\bea
(2.68)=\frac{8}{k^+ k^-}  \frac{  - k_{\perp}^2 q^2 + q_1^2 {q'_2}^2 +
 {q'_1}^2 q_2^2}{k^2} 
\eea
 An analogous result holds for the crossed graph. 

It is interesting to compare this result with (\ref{BFKL1prime1}): the use of the Ward identities has led to the disappearance of the term 
proportional to $q^2$. This was the term for which, in (\ref{BFKL1prime1}), 
the integral of the longitudinal momentum was divergent; furthermore, this term had 
prevented the graph to have the 'zero property'. It was only after taking the 
sum of both graphs, the uncrossed and crossed ones, that this term disappeared.
Now, after the use of the Ward identities, the longitudinal integration converges 
for each graph separately, and also the  zero property holds for each graph. 
With the help of the alternative expressions given in (\ref{BFKL1prime2}), this feature is seen most explicitly.
  
This argument shows that our use of the Ward identities is 
nothing but a rearrangement in the sum of the two terms, 
uncrossed plus crossed graph: in the representation (\ref{BFKL1}) 
each single graph has a divergent $k^-$ behavior, both in the 
infrared and ultraviolet region (the latter cancels in the sum). 
In contrast to this, in (\ref{BFKL3}) each term has a 'good' 
ultraviolet behavior. Our experience from using either of the 
two representations can be summarized by stating that both 
uncrossed and crossed graph contribute with equal weight.

We finally mention that (\ref{BFKL3}) could also have been obtained by proceeding in a 
different order: e.g., first the Ward identity  in $q'_1$, then in $q_1$, in $q_2$ ,and finally in $q'_2$ . Here the quartic coupling would have contributed also in intermediate steps.   

\section{General strategy}

After having presented this explicit calculation let us return to the identities formulated in 
section 1.

\subsection{Interpretation of BFKL results}
Let us first show that our results for the BFKL kernel, in fact, are equivalent to the 
Ward identities of section 1.

It will be convenient to introduce a few graphical notations.
Returning to the production vertex (\ref{prodvertex}) we introduce the following notation:
  \begin{center}
\epsfig{file=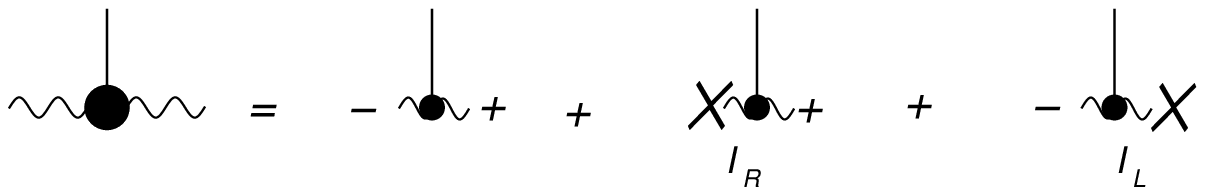,width=12cm,height=2cm}\\
Fig.2: induced vertices in the production vertex.
\end{center}
Straight lines stand for elementary gluons (not necessarily on shell), wavy lines for reggeized 
gluons. Lines to the left of the production vertex carry the polarization vector $(n^-)$, those to the right $(n^+)$.  The crosses denote the reggeons which end on an induced vertex (i.e. couple to two or more elementary gluons). Lines with a such a cross, in contrast to the reggeon lines without crosses, do not participate in the Ward identity. 
The blobs in the center denote all diagrams derived from the effective action (including all permutation of the external legs). 
Note that the quartic couplings only appear in the first diagram on the left.  
With this notation, we write symbolically in Fig. 3 the 4-point function as a sum of 11 terms:\\ 
  \begin{center}
\epsfig{file=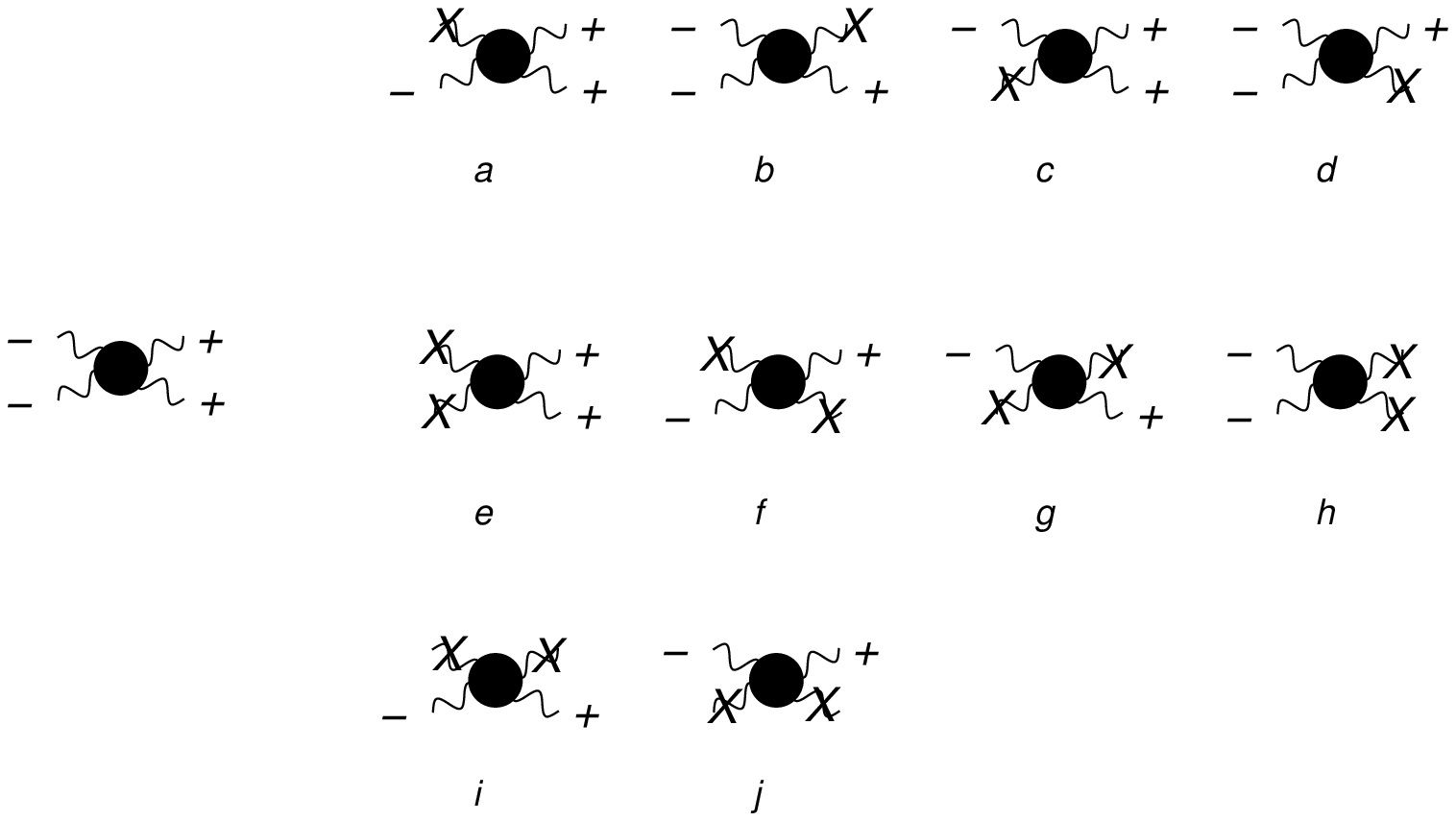,width=14cm,height=5cm}\\
Fig.3: induced vertices in the BFKL kernel.
\end{center}
Diagrams a-j denote the 10  contributions containing induced terms in one or two 
external legs.
 
 Now it is easy to summarize our findings.
 First we have shown the Ward identity in $q'_1$:  
 \be
 n^- n^- M^{(1)} q'_1 n^+=0. 
 \label{genWard1}
 \ee
where $M^{(1)}$ contains all terms except for 'b', 'g', 'h', and 'i' (for the remaining induced terms
it is understood that the contraction with the polarization vectors $n^+$, $n^-$ only applies to 
the legs without crosses; for example, in 'a' only the three legs with momenta $q'_1$, $q'_2$ and $q_2$ can be contracted). 
We then used this result to replace $(n^+)$ by $T(q'_1)= -\frac{2q'_{1\perp}}{{q'_1}^-}$. In short,
 \bea
  n^- n^- M^{(1)} n^+ n^+ = n^- n^- M^{(1)}\,T(q'_1)\, n^+.
\label{BFKL1a}
 \eea
 In order to arrive at the complete 4-point function, we have to re-add the induced terms 
 'b', 'g', 'h' and 'i'. Fig.3 illustrates  this result: the full 4-point function is given by the sum 
 of all diagrams where, for the upper right reggeon, the '+' is replaced by $T(q'_1)$.  The diagrams  
 'b', 'g', 'h', and 'i' remain unchanged. Note that, inside  these 10 induced terms, several 
 contributions actually vanish: this is because the induced vertices $I_L$ and $I_R$ vanish 
 when contracted with a purely transverse vector (see our comment after (2.36)). 
 
 In the second step, we have  used the Ward identity in $q'_2$:
 \be
 n^- n^- M^{(2)}\,T(q'_1)\,q'_2 =0,
 \label{genWard2}
 \ee
 where $M^{(2)}$ now contains all graphs except for those with a cross in the lower right reggeon 
('d', 'f', 'h', and 'j'). This identity allowed us to write  the 4-point function as sum of all diagrams of Fig.3, 
with the  '+'  in both the upper and the lower  reggeons on the rhs being replaced by 'T' (again, many contributions are actually zero). 
 
 Our third step derived the Ward identity in $q_1$, where all diagrams except for 'a', 'e', 'f', 'i' 
 contributed
 \be
 q_1 n^- M^{(3)} T(q'_1)T(q'_2) = 0,
\label{genWard3}
 \ee
  and we obtained a representation for the 4-point function where, in Fig.3, all 
 '-'  labels for the upper left reggeon are replaced by a'T'. Finally, the Ward identity in $q_2$:
 \be
 T(q_1) q_2 M^{(4)} T(q'_1)T(q'_2) =0 
 \label{genWard4}
 \ee
 led to our final representation of the 4-point function in which we have transverse momentum vectors for all reggeon lines
\be
T(q_1) T(q_2) M_0 T(q'_1) T(q'_2).
\ee
 A closer look at the induced terms shows that they are all absent, leading to the result in which only the first graph in Fig.3, $M_0$, contributes.
  
By combining the identities (\ref{genWard1}),  (\ref{genWard2}), (\ref{genWard3}), and (\ref{genWard4}) (plus others obtained by suitable permutations of the arguments) one derives 
the generalized Ward identities of section 1. For example, we interchange, in (\ref{genWard1}), the 
arguments on the rhs:
\be
 n^- n^- M^{(1)} n^+ q'_2 =0
 \label{genWard1a}
 \ee
(with a suitably modified $M^{(1)})$.
Taking the difference between (\ref{genWard1a}) and (\ref{genWard2}) we obtain:
\be
n^- n^- \tilde M q'_1 q'_2 = 0.
\label{genWard5}
\ee
Note that in $\tilde M$ the induced graphs 'b', 'g', and 'i' have cancelled. 
In a similar way one derives identities with three and four contractions, e.g. 
\be
q_1 q_2 M^{(4)} q'_1 q'_2 = 0.
\label{genWard7}
\ee

The identities (\ref{genWard1}), (\ref{genWard5}), and (\ref{genWard7}) are special cases of the general class of identities which have been discussed in 
section 1.
 
\subsection{A general strategy}

After this example it should have become clear how to use, in more general cases, the 
Ward identities for replacing unphysical polarization vectors by transverse momentum vectors.  
We simply invert the order of the arguments:  starting from the whole tower of Ward identities (\ref{EA_id}) and forming suitable linear combinations we arrive at our results  \ref{BFKL2} and \ref{BFKL3}.
Let us demonstrate this in detail:
\bea
q_1n^{-} M n^{+} n^{+}=0\nonumber\\
n^{-}n^{-} M q'_1  n^{+} =0\nonumber\\
q_1q_2 M n^{+}  n^{+} =0\nonumber\\
n^{-}n^{-} M q'_1  q'_2 =0\nonumber\\
q_1 n^{-} M q'_1  n^{+} =0\nonumber\\
q_1q_2 M q'_1 n^{+} =0\nonumber\\
q_1n^{-} M q'_1  q'_2 =0\nonumber\\
q_1q_2 M q'_1  q'_2 =0,
\label{rel4}
\eea
where, in each of these equations, $M$ contains a slightly different set of induced graphs (for simplicity, we always use the same symbol '$M$').   
In order to keep track of these induced terms, it is helpful to draw explicitly diagrams 
as we did in Fig.3. In the notation of section 1, the sum of all diagrams belongs to Green's functions with $A$ fields. In order to obtain, in  one of the identities (\ref{rel4}), a '$q$' contraction, one  
substitutes for the corresponding external leg $A \to \del_{\mu} v^{\mu}$. This  automatically removes all induced 
diagrams with a cross in the corresponding reggeon line, and the Ward identity applies to the sum of all remaining diagrams. In this way can easily see, for each of the equations 
(\ref{rel4}), which induced graphs are contained in $M$ and which ones are left out.   

Next we take linear combinations. For example, from the second equation of (\ref{rel4}),together with the identity 
\be
\frac{1}{{q'_1}^-} q'_1 = n^+ - T(q'_1)
\label{T-}
\ee
we derive 
\be
n^{-}  n^{-}  M n^{+} n^{+} =n^{-}  n^{-} M T(q'_1)n^{+}, 
\label{part1}
\ee
which coincides with our first result (after (\ref{BFKL1a})). The $M$ on the lhs contains all diagrams of Fig.3: in $\tilde M$ on the rhs all diagrams without  a cross in the reggeon $q'_1$ carry a '$T$' vector, the ones with a cross are unchanged. This defines 
$M$ on the rhs.  

Considering the second and fourth relations in eq.~\eqref{rel4} we can obtain another useful identity
\be
n^{-}n^{-}M q'_1  T(q'_2)  =0,
\label{rela1}
\ee
where all diagrams with a cross in the upper right reggeon are left out.
On applying this to the rhs of eq.~\ref{part1} we obtain our result (\ref{BFKL2}):
\be
n^{-}  n^{-}  M n^{+} n^{+} =n^{-}  n^{-}  M T(q'_1) T(q'_2), 
\label{part2}
\ee
where on the rhs of this equation all diagrams without crosses in $q'_1$ or $q'_2$  are contracted with 'T' vectors (some of them vanish). Diagrams with crosses remain unchanged.  

Let us construct some other relations (from now on we always use the same notation $M$ without 
mentioning in detail which diagrams are to be dropped of kept):
\bea
n^{-} q_2 M n^{+} q'_2  =0 , n^{-} q_2M  n^{+} n^{+} =0 \Rightarrow n^{-} q_2M  n^{+} T(q'_2)   =0 \nonumber \\
n^{-} q_2 M  q'_1 n^{+}  =0 , n^{-} q_2 M q'_1 q'_2  =0 \Rightarrow n^{-} q_2 M q'_1 T(q'_2)  =0 \,.
\eea
Combining the two results we arrive at 
\be
n^{-} q_2 M T(q'_1) T(q'_2)  =0.
\label{rela2}
\ee
Together with eq.~\eqref{part2} we find
\be
n^{-}  n^{-} M  n^{+} n^{+} =n^{-}  T(q_2) M  T(q'_1)T(q'_2).
\label{part3}
\ee
where $T(q_2)= -\frac{2q_{2 \perp}}{q_2^+}$, and we have defined, in analogy with (\ref{T-})
\be
\frac{1}{{q_2}^+} q_2 = n^- - T(q_2).
\ee 
Let us finally derive  our last result, (\ref{BFKL3}). From (\ref{rel4}) we derive 
\bea
q_1 q_2 M n^{+} q'_2  =0 , q_1 q_2 M n^{+} n^{+}  =0 \Rightarrow q_1 q_2 M n^{+} T(q'_2)  =0 \nonumber \\
q_1 q_2 M q'_1 n^{+} =0 , q_1q_2 M q'_1 q'_2  =0 \Rightarrow q_1 q_2M q'_1 T(q'_2)  =0 \,,
\eea
which can be combined into
\be
q_1 q_2 M T(q'_1)  T(q'_2)  =0.
\label{rela3}
\ee
We have also a relation similar to the one in eq.~\eqref{rela2}
\be
q_1 n^{-} M T(q'_1) T(q'_2)  =0.
\label{rela4}
\ee
Combining the last two equations  we obtain
\be
q_1 T(q'_2)  MT(q'_1) T(q'_2)  =0.
\label{rela5}
\ee
Finally, combining this  with eq.~\eqref{part3}, we get our final result for the BFKL kernel (\ref{BFKL3}):  
\be
n^{-}  n^{-} M  n^{+} n^{+} =T(q_1)   T(q_2)  M  T(q'_1) T(q'_2). 
\label{part4}
\ee     

In a future investigation we will apply this technique to the $3 \to 3$ Green's function, deriving the 
useful relation
\be 
n^+ n^+ n^+ M n^- n^- n^- = T(q_1) T(q_2) T(q_3) M  T(q'_1) T(q'_2) T(q'_3).
\ee

\section{Outlook}
In this paper we have derived Ward identities of Green's functions and scattering amplitudes 
involving physical (on-shell) particles and reggeized gluons. As a first application, we have verified 
these identities for the BFKL kernel, and we have arrived at a new representation.

We believe that these identities will be particularly useful for computing higher order Green's functions of  reggeized gluons, e.g. $n \to m$ Green's functions. The simplest case, the BFKL Green's, has been originally derived from $s$-channel unitarity, i.e from the  imaginary part 
of a $2 \to 2$ scattering amplitude. In such a derivation, the produced 
$s$-channel gluon is taken to be on shell, and the corresponding $\delta$-function can be used to perform the integration over one of the longitudinal components of the momentum variables.
However, in higher order Green's functions, e.g. in the $3 \to 3$ case of the Odderon, the $3 \to3 $ 
kernel can no longer be derived from $s$-channel discontinuities, and a full integration over 
longitudinal momenta of $s$-channel gluons has to be performed. This raises the question of 
the convergence of the integration, which is obtained only after summing over all the permutations needed to have Bose symmetry.
The Ward identities discussed in this paper can be used to obtain ultraviolet convergent expressions of single contributions without the need to perform the full sum. 
In the context of our discussion of the BFKL kernel we have demonstrated that is is useful to have at hand a representation in which the integration over longitudinal momenta is convergent in the 
ultraviolet region: this will become crucial to simplify the calculations for the case of the LO $3\to 3$ kernel which is currently under investigation \cite{BFLV}.

From the general point of view these Ward identities permit to rewrite $n\longrightarrow m$ reggeized gluon transition amplitudes (as in (\ref{BFKL3})) 
without induced terms. This suggests that one may search for a formulation of the Effective Action, equivalent to the existing one, in which the induced terms can be completely omitted.
A similar approach  can be also repeated in gravity in an Effective Action which includes reggeized gravitons.
\\[1cm]
{\bf Acknowledgements:} \\
J.B. gratefully acknowledges the support of the Galileo Galilei Institute where part of this work was done.
He has also been supported by the Premio 2011 de Excelencia Cientifica 'Abate Juan Ignacio Molina'. 
G.P.V. thanks the II Institut f\"ur Theoretische Pkysik of the Hamburg University for the hospitality. 


\end{document}